\documentclass[%
 aip,
 amsmath,amssymb,
 reprint,%
]{revtex4-1}

\usepackage{graphicx}
\usepackage{dcolumn}
\usepackage{bm}
\usepackage{float}

\usepackage[utf8]{inputenc}
\usepackage[T1]{fontenc}
\usepackage{mathptmx}

\begin{document}

\preprint{AIP/123-QED}

\title{Microwave Engineering for Semiconductor Quantum Dots in a cQED Architecture}

\author{Nathan Holman}
\affiliation{Department of Physics, University of Wisconsin-Madison, 53703}

\author{J.P. Dodson}
\affiliation{Department of Physics, University of Wisconsin-Madison, 53703}

\author{L.F. Edge}%
\affiliation{HRL Laboratories LLC, 3011 Malibu Canyon Road, Malibu, CA 90265, USA}

\author{S.N. Coppersmith}%
\affiliation{Department of Physics, University of Wisconsin-Madison, 53703}
\affiliation{University of New South Wales, Sydney, Australia}

\author{M. Friesen}%
\affiliation{Department of Physics, University of Wisconsin-Madison, 53703}

\author{R. McDermott}%
\affiliation{Department of Physics, University of Wisconsin-Madison, 53703}

\author{M.A. Eriksson}
\affiliation{Department of Physics, University of Wisconsin-Madison, 53703}

\date{\today}

\begin{abstract}
We develop an engineered microwave environment for coupling high \emph{Q} superconducting resonators to quantum dots using a multilayer fabrication stack for the dot control wiring. Analytic and numerical models are presented to understand how parasitic capacitive coupling to the dot bias leads can result in microwave energy leakage and low resonator quality factors. We show that by controlling the characteristic impedance of the dot bias wiring, on-chip quality factors of $8140$ can be attained without the addition of explicit filtering. Using this approach we demonstrate single electron occupation in double and triple dots detected via dipole or quadrupole coupling to a superconducting resonator. Additionally, by using multilayer fabrication we are able to improve ground plane integrity and keep microwave crosstalk below -20~dB out to 18~GHz while maintaining high wire density which will be necessary for future circuit quantum electrodyanmics (cQED) quantum dot processors. 
\end{abstract}

\maketitle

Gate defined quantum dots are a nascent platform for quantum computing in which electron charge and spin states are used to define the quantum bit\cite{Hanson,Zwanenburg}. In silicon and Si/SiGe heterostructures, recent work has shown it is possible to fabricate single, two, and four qubit systems\cite{yoneda,dzurak,Kawakami:2016p11738,Zajac3,Sigillito,Mi_2,watson} with control infidelities at or approaching the $10^{-2}$ level requisite for error correction\cite{Fowler}. For most quantum dot circuits, the control wiring scheme consists of an electron beam (e-beam) defined gate electrode structure with a rapid fan out into pads that are connected by aluminum or gold wire bonds to a printed circuit board (PCB) with typical die sizes of only a few millimeters. While this keeps the fabrication complexity to a minimum, it results in a minimally controlled microwave environment for the device and any readout circuitry. Near term quantum computing efforts with quantum dots face significant quantum systems engineering challenges balancing the needs for high fidelity readout, coupling, and control.
\par In this letter, we demonstrate a wiring scheme for quantum dot devices in a cQED architecture. This approach allows for high-density, low crosstalk wiring with controlled RF leakage characteristics, paving the way for larger quantum dot processors utilizing cQED techniques \cite{Mi_2,Stockklauser,Samkharadze}. A simple and intuitive circuit model for cavity leakage from external leads reveals that minimization of the impedance of the environment at the cavity frequency ameliorates photon leakage out the gate leads, which otherwise present an undesired load on the cavity. We implement this method with a microstrip wiring scheme that achieves a low characteristic lead impedance of $Z_g \approx 10$~$\Omega$, and demonstrate resonators with quality factors as high as 8140 (design $Q_L = 10^4$) while connected to the quantum dot gate stack.
\begin{figure}[ht!]
\includegraphics[width=\columnwidth]{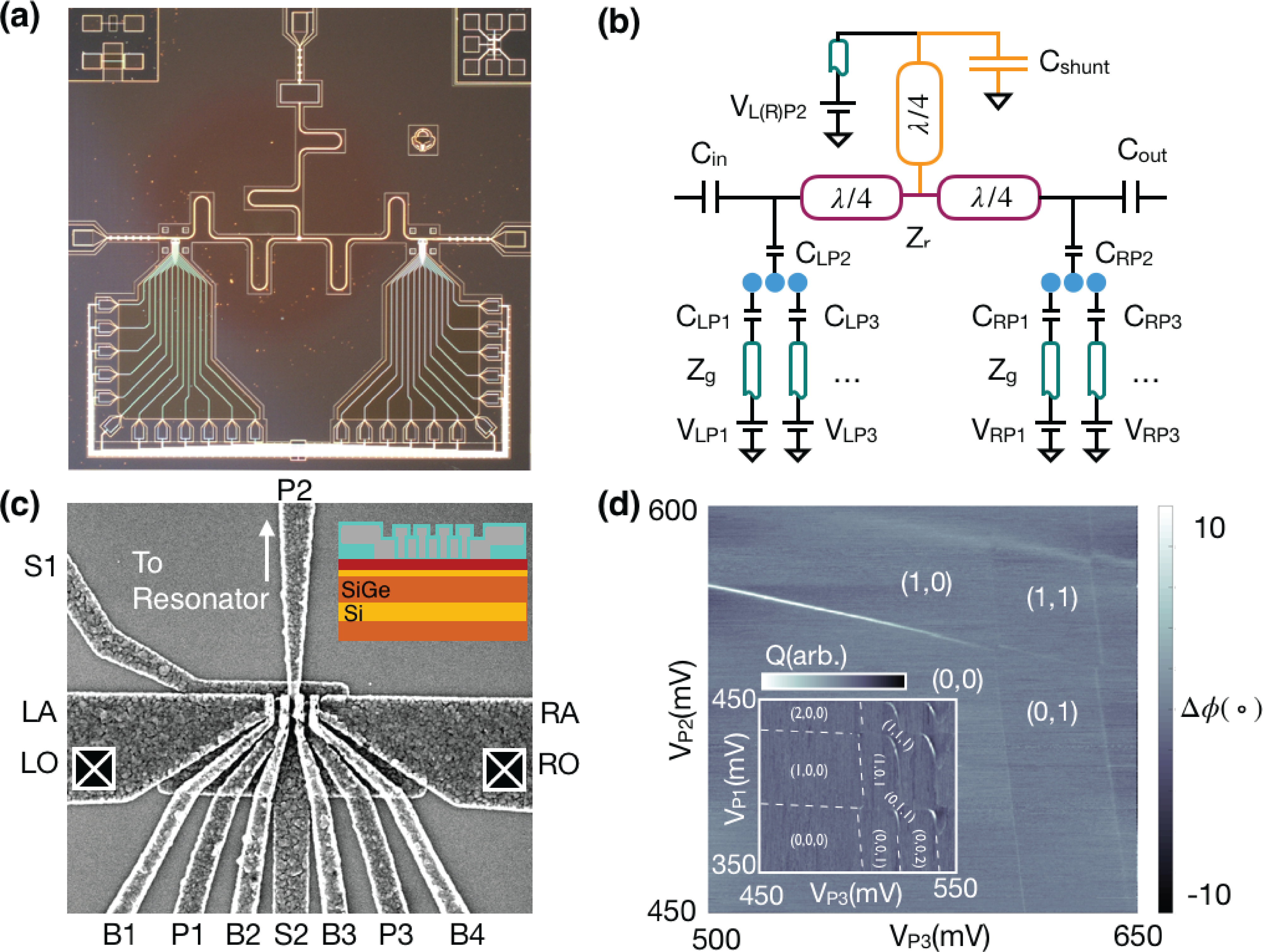}
\caption{\label{fig:test1} (a) Dark field optical micrograph of a device using microstrips for the quantum dot leads (lower half of die). (b) A simplified circuit model for the device. The cavity (purple) is a half-wave superconducting microwave resonator with a quarter-wave impedance transformer with shunt capacitance (orange) connected to its voltage node. Transmission lines with characteristic impedance $Z_g$ (teal) are used to supply bias voltages to tune the dots. (c) Scanning electron microscope image of a typical triple dot device. The center dot plunger pin is galvanically tied to the resonator allowing for microwave detection of both double and triple quantum dots. Inset: a schematic cross section along the active region of the device. (d) Cavity phase detection of a double quantum dot stability diagram in the single-electron regime under P2 and P3. A linear background was subtracted from the raw data for clarity. Inset: Cavity detection of a triple dot stability diagram in the single-electron regime.
}
\end{figure}
\par For these experiments, we fabricate cavity-coupled pairs of triple quantum dot structures on an undoped Si/SiGe heterostructure using an overlapping aluminum gate stack\cite{zajac1}. The two dimensional electron gas (2DEG) is formed in a 9~nm thick strained silicon layer either 20 or 30~nm below the surface. To avoid accumulation of 2DEG under the resonator, we remove the heterostructure everywhere except for two 50~x~100~$\mathrm{\mu m}^2$ mesa regions on the sample where the dot structures are located. The circuit consists of four key components: the overlapping aluminum gate electrodes, the dot lead wiring, the superconducting $\lambda/2$ microwave resonator, and the resonator DC voltage tap. An optical image of a finished device is shown in Fig. \ref{fig:test1} (a). To understand the control environment of the device, Fig. \ref{fig:test1} (b) provides a simplified circuit diagram illustrating the key components  the on-chip microwave engineering. Additional wires and parasitic cross capacitances are not drawn for simplicity. We confine the 2DEG into quantum dots using three layers of overlapping aluminum gates patterned with e-beam lithography (Fig. \ref{fig:test1} (c)) and deposited by e-beam evaporation. High yield electrical isolation between gate layers is achieved by cleaning and oxidizing the aluminum after lift-off of each layer with a 250 W downstream oxygen plasma asher for 10 minutes\cite{dodson}.
\par Measurements of these samples are performed by wire bonding each device in a hybrid PCB-metal box enclosure designed to raise unwanted chip-mode frequencies to >20~GHz by creating a milled pocket below most of the 6.15~x~6.15~$\mathrm{mm}^2$ die \cite{Wenner}. The packaged sample is cooled in a dilution refrigerator with base a temperature of $T_\mathrm{mc} = 25  \text{ mK}$ and typical electron temperatures between 80-100~mK determined by fitting to a thermally broadened conductance peak\cite{Beenakker}. Charge detection is achieved by measuring the cavity-charge interaction during electron tunneling events\cite{Frey1,Mi_1}. These interactions are formed by connecting a gate (in our case P2 in Fig. \ref{fig:test1} (c)) to an aluminum or niobium microwave $\lambda/2$ resonator. Zero point fluctuations in the electric potential of the \emph{LC} oscillator couple to the dipolar (quadrupolar) detuning degree of freedom ($\epsilon$) for the charge in the double (triple) dot system, allowing for detection of the quantum capacitance $ C_Q \propto \frac{d^2E_{01}}{d\varepsilon^2}$. By probing the cavity with a microwave tone at the bare cavity frequency ($f_r$) and recording the transmitted amplitude and phase as a function of plunger gate voltage, damping and phase shifts of the probe are observed during electron tunneling events when the rates are comparable to the cavity frequency (of order several GHz). In the overlapping gate architecture, this condition is easily achieved through tuning of the dot barrier gate voltages B1-B4.  Using the plunger gates P1-P3, we can empty out the electrons in a double (or triple) dot and reach the (0,0) (or (0,0,0)) charge state configuration, as demonstrated in Fig. \ref{fig:test1} (d).
\par In order to maintain a high cavity quality factor we engineer the microwave environment to minimize photon leakage out the 25 bias leads of the two triple dot gate structures. This engineering amounts to maximizing the reflection coefficient of the microwave energy out any lead other than the readout port by making the other leads look like high or low impedance to ground. A unique requirement for cQED experiments with dots is DC voltage biasing of the center pin of the coplanar waveguide (CPW) resonator in order to accumulate a quantum dot under P2. Test devices with DC taps at the center of the waveguide had quality factors less than $10^3$, which was much lower than the explicit coupling defined by $C_{\text{in}}$ and $C_{\text{out}}$. We eliminate leakage out this lead by turning the tap into a $\lambda/4$-length CPW. The end of the $\lambda/4$ CPW is shunted with a load impedance $Z_{L}$ in the form of a large parallel plate capacitor using an $\text{SiO}_2$ dielectric. The input impedance for a $\lambda/4$ segment of transmission line terminated by a load $Z_L$ is given by\cite{Pozar}
\begin{equation}\label{eqn:quarterwave}
    Z_{\text{in}} = Z_0 \frac{Z_L + i Z_0 \tan\left(\beta\ell\right)}{Z_0 + i Z_L \tan\left(\beta\ell\right)}\biggr\rvert_{\ell = \lambda/4} =  \frac{Z^{2}_{0}}{Z_{L}},
\end{equation}
where $\beta = 2\pi/\lambda$ is the propagation constant of the transmission line, and $\ell$ is its physical length. By making the shunt capacitance large ($C_{\text{shunt}}\approx100$~pF), we achieve $Z_{in} \approx 10 \text{ k}\Omega$. In combination with tapping at the voltage node, this effective impedance leads to minimal leakage out the DC bias tap at the resonance frequency. Using this DC bias, devices without the overlapping aluminum gates achieved loaded quality factors as high as $4\text{x}10^4$, well beyond the limit imposed by parasitic loading from the quantum dot circuit. We note that the quadratic dependence of $Z_\mathrm{in}$ on $Z_0$ provides a way to further increase the quarter-wave tap input impedance through use of large $Z_0$ CPWs\cite{Samkharadze}.
\par The use of the overlapping gate stack poses a unique challenge for RF readout schemes, because of the large parasitic capacitances $C_p$ (order 1~fF) between the gate electrodes in the region where the dots are formed. These parasitic capacitances are the same order of magnitude as the capacitance used to purposefully couple photons into the readout port ($C_{\text{out}} \approx 5.5$~fF for a quality factor of $10^4$ at 7.25~GHz) and therefore result in substantial microwave leakage out the leads. To analyze this leakage we first use a lumped element circuit model for loading of an \emph{LCR} oscillator to a transmission line environment with characteristic impedance $Z_g$ and parasitic capacitance $C_p$ \cite{Pozar,Goppl}. We then extend this model to real circuits with finite length leads where the general transmission line impedance transformer behavior captured in Eq.~(\ref{eqn:quarterwave}) must be taken into account. After including finite length effects we show that the intuition from the lumped element model is qualitatively correct as long as dot gate lead lengths sufficiently avoid half integer multiples of the resonance wavelength $\lambda$ (e.g. $n\lambda/2$ for $n=0,1,2...$).
\par
\begin{figure}[ht]
\includegraphics[width=1\columnwidth]{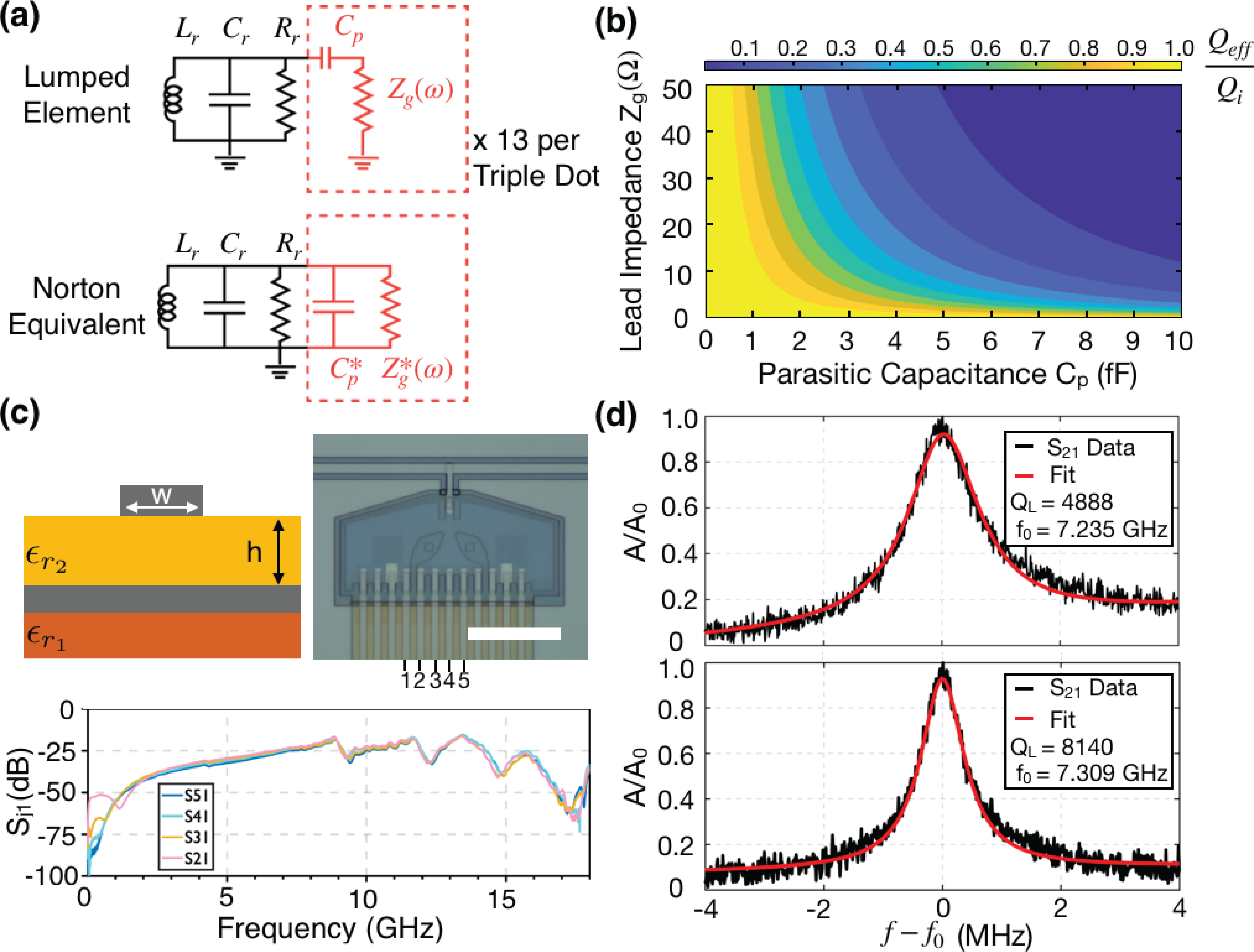}
\caption{\label{fig:test2} (a) Lumped element model for an \emph{LCR} oscillator coupled to a parasitic lead and its Norton equivalent circuit. (b) A contour plot of $Q_{\text{eff}}/Q_i$ as a function of gate impedance and parasitic capacitance assuming $Q_i = 10^5$. (c) A schematic of a microstrip line consisting of a substrate ($\epsilon_{r1}$), ground plane, pad dielectric ($\epsilon_{r_2}$) of thickness $h$, and strip of width $W$. Adjacent is an optical image of a mesa structure with microstrip leads on a 7~$\mu$m pitch (scale = 50~$\mu\text{m}$). Below: the microwave crosstalk between numbered microstrips connected to the set of gates used for biasing (P1:B2:S2:B3:P3) indicating -20~dB crosstalk between leads at frequencies above 5~GHz for a 7~$\mu$m pitch. Below 3~GHz, crosstalk is dominated by proximity of bond pads rather than proximity of the microstrips. Ripple features in the crosstalk are due to the finite lengths of the microstrips and their mismatch to the 50~$\Omega$ coaxial cables. (d) Example resonator transmission spectra from two devices with $Q_L = 4888$ and $Q_L = 8140$.}
\end{figure}

 For the lumped element analysis, we write the resonator intrinsic quality factor as $Q_i = \omega_{\text{\emph{LC}}}\text{R}_r C_{r}$. Here $R_r$ is the effective damping resistance arising from the lossy dielectrics coupled to the electric field of the resonator. Schematically, each dot lead looks like the series combination of a capacitance $C_p$ and, in the limit of an infinite transmission line, an effective real impedance $Z_g$ to ground (Fig. \ref{fig:test2} (a)). In the Norton equivalent circuit the gate impedance transforms to an additional parallel resistance
\begin{equation}
    Z_g^{*} = Z_g \left(1+\frac{1}{\omega^2C_p^2Z_g^2}\right).
\end{equation}
By computing the new total load resistance $R_{||}=R_r||Z_g^{*}$, we find the effective quality factor $Q_{\text{eff}} = \omega_{\text{\emph{LC}}}\text{R}_{||}C_{r}$. In terms of the impedance of the parasitic capacitor $Z_c$, gate impedance $Z_g$, and effective internal resistance $R_r$ we find:
\begin{equation}\label{eqn:Qeff}
    \frac{Q_{\text{eff}}}{Q_i} = \frac{Z_c^2 + Z_g^2}{Z_c^2 + Z_g^2\left(1 + \frac{R_r}{Z_g}\right)},
\end{equation}
revealing there are in principle two ways for reducing the effect of unwanted loading: minimize the parasitic capacitance (maximizing $Z_c$) or alter the gate impedance $Z_g$ to minimize the effect of the $R_r/Z_g$ term in the denominator of Eq.~(\ref{eqn:Qeff}). Notably, it is easier to reach the limit of $Z_g \ll R_r$ than $Z_g \gg R_r$, as $R_r\gtrsim 10$ $\text{M}\Omega$ for typical superconducting resonators. Figure \ref{fig:test2} (b) shows a contour plot of Eq.~(\ref{eqn:Qeff}) in the low $Z_g$ regime. For our experiments, the characteristic cavity impedance is $Z_r = 50$ $\Omega$, but the core result holds for higher impedance cavities as well and is captured by the definition of $R_r$.
\par Since reducing the parasitic capacitance in the overlapping gate stack is intrinsically difficult, we opt to control the gate impedance $Z_g$. Previous efforts to improve cavity quality factors involve the use of RF choke inductors or \emph{LC} low pass filters, which have either high or low impedance at the cavity frequency, suppressing leakage\cite{Frey1,Mi_1}. In both cases, the leads acquire a frequency dependent filter function which substantially limits the control bandwidth. To eliminate this potentially undesired feature, we choose to reduce the characteristic impedance of the transmission line on-chip using a microstrip geometry. The design and implementation of microstrip wiring is illustrated in Fig. \ref{fig:test2} (c). The wires are fabricated in a multilayer fabrication process that has three essential steps: deposition of the base layer ground plane, growth of an insulating dielectric layer, and deposition of the microstrip counter electrode.
\par To obtain the desired low impedance, we fabricate microstrips with a width $W = 3$ $\mu\text{m}$ and an $\text{SiO}_2$  thickness $h = 0.2$ $\mu\text{m}$ yielding the limit where the parameter $\alpha = W/h \gg 1$. The impedance of the microstrip in this limit can be calculated using conformal mapping and is given by\cite{Gupta,Pozar}:
\begin{equation}
    Z_g \approx \frac{1}{\alpha+ 1.393 + 0.667\log(\alpha + 1.444)}\sqrt{\frac{\mu_0}{\epsilon_0\epsilon_e}}
\end{equation}
with an effective permittivity: 
\begin{equation}
    \epsilon_e \approx \frac{\epsilon_{r_2}+1}{2} + \frac{\epsilon_{r_2}-1}{2\sqrt{1+12/\alpha}}.
\end{equation}
For our parameters, we find $Z_g \approx 10~\Omega$ which compares favorably to \emph{LC} low pass filters in the literature\cite{Mi_1}, that have an input impedance of $Z_{\text{in}} \approx 20~\Omega$, while retaining the flexibility of a flat frequency response. The low impedance leads come at the cost of an insertion loss of approximately 3.5~dB due to mismatch between the $Z_0 = 50$ $\Omega$ coaxial cabling and the $Z_g = 10~\Omega$ microstrip. This effect could be mitigated by an impedance matching element such as a Klopfenstein taper\cite{Gupta}.
\par The multilayer stack also provides a means for improved and reliable microwave performance in increasingly complex processors through additional on-chip crossovers. These structures serve as low inductance connections between ground planes on-chip, thereby suppressing spurious slot line modes more efficiently than with traditional aluminum wire bond stitching \cite{Chen1}. Furthermore, aluminum wire bond stitching suffers from the low critical field of aluminum ($H_c \approx 10$~mT), which can damp the cavity upon application of an external magnetic field\cite{petersson}. Since both the crossover and base ground plane metals can be made of field tolerant superconductors such as niobium, they are useful for preserving circuit performance in cavity-spin coupling experiments.  Lastly, in Fig. \ref{fig:test2} (c) we demonstrate that this microstrip wiring scheme maintains less than -20~dB crosstalk between adjacent leads over a broad frequency range with a 7~$\mu$m wire pitch for approximately 1~mm, which could be important for minimizing off-resonant driving observed in recent resonant two-qubit gate experiments\cite{Zajac3, Sigillito}.
\begin{figure*}[t]
\includegraphics[width=2\columnwidth]{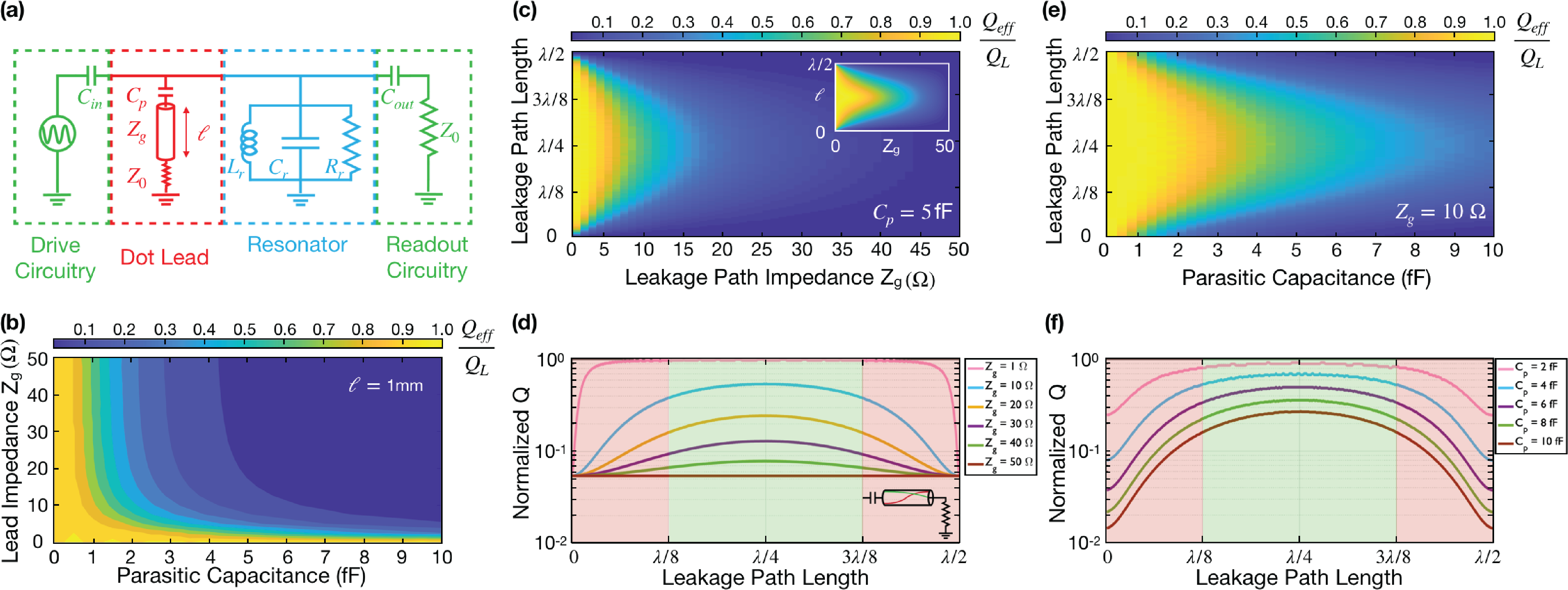}
\caption{\label{fig:test3} (a) Circuit schematic for the SPICE simulations of loading from an impedance-mismatched lead (red) of variable length to a 50~$\Omega$ environment. The resonator (blue) with explicit external loading from the drive and readout circuity (green) has a design $Q_L = 10^5$ and frequency $f_r = 7.25$~GHz. (b) A contour plot of  $Q_{\text{eff}}/Q_L$ for a 1~mm long leakage path as a function of its characteristic impedance and parasitic capacitance. (c) Color plot of $Q_{\text{eff}}/Q_L$ as a function of the leakage path length and impedance ($Z_g$). Inset: the reflection coefficient of a 50~$\Omega$ terminated transmission line computed using Eq.~(\ref{eqn:quarterwave}). (d) Line cuts from (c) for different $Z_g$ showing the impact on the $Q_{\text{eff}}/Q_L$ as a function of the lead length. Inset: the nominal voltage profiles for the quarter-wave (green) and half wave (red) lengths of transmission line. (e) Color plot of the $Q_{\text{eff}}/Q_L$ as a function of leakage path length and parasitic capacitance. (f) Line cuts from (e) for different $C_p$ as a function of lead length. In both (d) and (f) flattening in $Q_{\text{eff}}/Q_L$ is the result of the leakage becoming less than the design $Q_L = 10^5$.}
\end{figure*}
\par Using these design principles, we fabricated and measured over twenty cavity coupled quantum dot structures and found a total spread in resonance quality factors between 2600-8140 (example spectra shown in Fig. \ref{fig:test2} (d)). We attribute the spread in quality factor to variation in parasitic capacitance and finite length effects for the low impedance microstrips (discussed below). We note that in addition to the spread in quality factors for identical resonator designs (with lithographic variation of order 0.2~$\mu\text{m}$), the fundamental resonance frequencies show a large spread of approximately $\pm 100$~$\text{MHz}$. Using the measured frequency, length of the waveguide, and conformal mapping\cite{Gupta,Goppl} to estimate the capacitance of the CPW structure, we back out a mean substrate permittivity to be $\epsilon_\text{r}^{\text{SiGe}} \approx 14.2$. Using linear interpolation\cite{Schaffler} we estimate the permittivity of the SiGe alloy to be $\epsilon_\text{r}(x) \approx \epsilon_\text{r}^{\text{Si}}(1-x)+\epsilon_\text{r}^{\text{Ge}}x\approx 13$ where $x=0.3$. The origin of the apparent deviation is unknown at this time. Variation in the observed resonance frequencies could be explained by variation in the relative dielectric constant of the SiGe alloy by $\pm 0.3$, possibly caused by fabrication processing or growth variation across the wafer.
\par Although Eq.~(\ref{eqn:Qeff}) is correct\cite{Goppl,Pozar} for coupling to an infinitely long transmission line with impedance $Z_g$, in practice the low impedance leads on-chip are only a few millimeters long and are wire bonded to a $50$~$\Omega$ environment. To elucidate the impact of the finite length of on-chip low impedance leads, we use "Simulation Program with Integrated Circuit Emphasis" (SPICE) to calculate the resonator line width under various conditions. Figure \ref{fig:test3} (a) shows the circuit schematic used to understand the consequences of using finite transmission lines of length $\ell$. Similar to the lumped element case, lowering $Z_g$ and $C_p$ results in suppression of leakage. For a $\ell = 1\text{ mm}$ microstrip, the contour plot of the leakage path impedance in Fig. \ref{fig:test3} (b) is qualitatively similar to that in Fig. \ref{fig:test2} (b), but the roll-off is stronger as a function of $Z_\mathrm{g}$.
\par In Fig. \ref{fig:test3} (c,d) we compute $Q_\mathrm{eff}/Q_\mathrm{L}$ as a function of $Z_\mathrm{g}$ and leakage path (gate) length. When the gate length is a quarter wavelength of the resonance frequency, no voltage drop occurs across the $50$~$\Omega$ environment at the end of the lead, effectively eliminating loss out the lead. Conversely, when the lead is exactly $\lambda/2$ in length, the voltage drop across the $50$~$\Omega$ environment is maximal, resulting in loss to the environment equivalent to using no leakage suppression scheme at all. As shown in Fig. 3(c,d), if $Z_g$ is low, the range of gate lead lengths that suppress leakage becomes very wide, with only lengths very close to $\lambda/2$ displaying any significant degradation in $Q_\mathrm{eff}$. Figure~\ref{fig:test3} (e,f) shows that analogous results hold for $Q_\mathrm{eff}$ as a function of $C_p$. 
The inset to Fig.~\ref{fig:test3} (c) shows the reflection coefficient of a 50~$\Omega$ terminated transmission line calculated using Eq.~(\ref{eqn:quarterwave}).  The result is very similar to the SPICE calculation in the main panel, and thus the simple analytic form provides good intuition for the finite length effects in maintaining a high \emph{Q}.
\par In conclusion, we demonstrated a high-density, low-crosstalk, low impedance wiring scheme for quantum dots in a cQED framework. Using a simple circuit  model we designed a filterless low impedance wiring strategy for minimal microwave leakage achieving quality factors as high as 8140. We showed how this approach remains robust even in the presence of finite length effects of the low impedance leads. For devices with functional quantum dots, measurement of the charge configuration down to zero electrons in both double and triple dots is achieved, paving the way towards study of spin-photon coupling of exchange-based qubits in Si/SiGe \cite{Russ1}.
\\
\\
The data that support the findings of this study are available from the corresponding author upon reasonable request.
\par We acknowledge discussions on multilayer superconducting circuit fabrication with A. Opremcak, E. Leonard, M. Beck, F. Schlenker, and M. Vinje. We acknowledge technical advice in development of processing recipes by K. Kuptcho, Q. Leonard, and E. Gonzales. Research was sponsored in part by the Army Research Office (ARO) under Grant Numbers W911NF-17-1-0274 and by the Vannevar Bush Faculty Fellowship program under ONR grant number N00014-15-1-0029. We acknowledge the use of facilities supported by NSF through the UW-Madison MRSEC (DMR-1720415) and the MRI program (DMR–1625348). The views and conclusions contained in this document are those of the authors and should not be interpreted as representing the official policies, either expressed or implied, of the Army Research Office (ARO), or the U.S. Government. The U.S. Government is authorized to reproduce and distribute reprints for Government purposes notwithstanding any copyright notation herein.

\nocite{*}
\bibliography{aipsamp}

\end{document}